# Preparation, structure evolution, and magnetocaloric effects of MnFe(PGe) compounds

**Ming Yue**[1,a)], **Hongguo Zhang**[1], **Danmin Liu**[2], **and Jiuxing Zhang**[1]

1. College of Materials Science andEngineering, Beijing University of Technology, Beijing 100124, China
2. Institute of Microstructure and Property of Advanced Materials, Beijing University of Technology,Beijing 100124, China

a) Corresponding author, Email: yueming@bjut.edu.cn

**Abstract**

As one of magnetic refrigerants with giant magnetocaloric effect (GMCE), MnFePGe-based compounds had drawn tremendous attention due to their many advantages for practical applications. In this paper, correlations among preparation conditions, magnetic and crystal structures, and magnetocaloric effects (MCE) of the MnFePGe-based compounds are reviewed. Structure evolution and phase transformation in the compounds as a function of temperature, pressure, and magnetic field were reported. Influences of preparation conditions to the chemical composition and microstructure homogeneity of the compounds, which play key role to the MCE and thermal hysteresis of the compounds, were introduced. Based upon these experimental results, a new method to evaluate MCE of the compounds via DSC measurements was proposed. Moreover, the origin of "virgin effect" of the MnFePGe-based compounds was discussed.

1. Introduction

Magnetic refrigeration (MR) based on magneto-caloric effect (MCE) first reported by Warburg [1] in 1881 is a kind of advanced cooling technology. Up to now, different temperatures from below 1 K to near temperature (RT) had been successfully achieved by using adiabatic demagnetization of various magnetic refrigerants. In 1933, Giauque et al [2] obtained an ultralow temperature of below 1 K by MR technology with $Gd_2(SO_4)_3 \cdot 8H_2O$ as working medium. Later in 1976, Brown [3] applied the same technology to RT refrigeration and reached 80 K below RT by using Gd as refrigerants, opening an avenue to room temperature magnetic refrigeration (RTMR).

The main driving force for current research and development of RTMR is the requirement to energy efficient and environmentally friendly cooling technology, which is the major disadvantage of current vapor-compression refrigeration (VCR). RTMR technology bears higher cooling efficiency than VCR does, and it never uses harmful gas. Moreover, it also possesses other advantages such as high compaction, low noise, and high stability. Therefore, RTMR has draw tremendous attention from both scientists and engineers as a promising cooling technology for practical application. On the other hand, the entropy changes of magneto-caloric materials did not surpass that of Gd until the discovery of $Gd_5Si_2Ge_2$ compound by Pecharsky et al [6] in 1997. The compound undergoes a first-order magnetic transition in the vicinity of their Curie Temperature ($T_C$), leading to giant MCE (GMCE) and therefore remarkably larger entropy change that of Gd metal. Afterwards several new magneto-caloric materials with GMCE such as $La_{1-x}Ca_xMnO_3$[4], $LaFe_{13-x}Si_x$[5-7], $MnAs_{1-x}Sb_x$[8], and $MnFeP_{1-x}As_x$[9,10] had been developed. These achievements promote the research on RTMR up to a new level.

In 2002, Tegus et al [12] reported GMCE in $MnFeP_{1-x}As_x$ compounds with tunable $T_C$ via changing P/As ratio. In subsequent research, it is surprisingly found that the GMCE can be well preserved in similar MnFe(P, Ge) compounds, in which the toxic element As was replaced by Ge for ease of practical application [11-13]. However, the new MnFe(P, Ge) compounds also present some undesirable behaviors such as anomalous "virgin effect" and large thermal hysteresis. Fortunately, intensive investigations on the crystal/magnetic structure and phase transformation of the compounds have shed more light on the mechanism of the unexpected behaviors, and some effective ways have been proposed to improve the MCE and restrain the thermal hysteresis simultaneously. Moreover, several methods have been developed for properly evaluating the MCE of the compounds based on the gradually increased understanding of the combined structural and magnetic phase transition of the compounds.

In present paper, we give a brief review of the recent advances in MnFe(PGe) compounds, mainly on our recent progress in preparation, structure evolution, and magneto-caloric effects in

MnFe(PGe) compounds. In section 2 of this paper we will discuss preparation of the MnFe(PGe) compounds with spark plasma sintering (SPS) technology. Section 3 is devoted to an overview of temperature, magnetic field, and pressure dependence of crystal and magnetic structures evolution as well as phase transformation in MnFe(PGe) compounds based mainly on neutron powder diffraction (NPD) investigations. Section 4 describes an easy and reliable method for evaluating the MCE of the compounds via differential scanning calorimeter (DSC) technology. Section 5 talks about the behavior and origin of the "virgin effect" in MnFe(PGe) compounds based on magnetic measurements and Mössbauer spectroscopy study. In section 6 a brief conclusion of the entire paper will be driven.

## 2. Preparation of MnFe(PGe) compounds

In previous works, the MnFe(PGe) compounds had been prepared either by a mechanically activated solid-diffusion method[12,14] or by melt-spun method[15]. However, both methods require very long processing time and frequently involve the formation of undesirable inter-metallic ferromagnetic impurities in the compound. Such disadvantages potentially prevent the compounds from practical application.

SPS is an advanced consolidation technique that can produce materials under non-equilibrium conditions[16]. One of the important advantages of the SPS technique is the plasma-aid sintering mechanism, which effectively shortens the diffusion paths to ease phase formation and homogenization of the compound. In addition, the high sintering speed of SPS can restrain the phase separation and allow the consolidation of compounds in a short time. In our works, SPS has been applied to prepare bulk MnFe(PGe) compounds.

Yue et al[17] reported preparation of $Mn_{1.1}Fe_{0.9}P_{0.8}Ge_{0.2}$ compound by simple blending and subsequent SPS method. The high purity starting materials, Mn powders, Fe powders, Ge chips, and red P powders, were put together and manually blended under Ar atmosphere in the glove box. The as-blended powders were then collected into a carbon mold and fast consolidated into a cylindrical sample at 1193 K under 30 MPa by the SPS technique. The density of the sample was examined by the Archimedes method to be over 95% of the density of the as-cast same composition ingot. Fig. 1(a) shows the observed and calculated XRD patterns for the $Mn_{1.1}Fe_{0.9}P_{0.8}Ge_{0.2}$ compound. Apart from the $Fe_2P$ type main phase, a minor impurity phase of MnO, which probably originated from the starting materials or from the oxidation of Mn during the preparation process, accounts for less than 2 *wt.* % in the sample according to the refinement results. It is worth mentioning that some ferromagnetic secondary phase such as $(Mn, Fe)_5Ge_3$, which persists in conventionally sintered and cast samples even after long time annealing[12,14,15], are not presents in the present bulk sample prepared by the SPS technique. Magnetic measurement indicates that the $T_C$ of the compound is at 253 K and the thermal hysteresis is 15 K. In addition, the compound possesses the maximum magnetic entropy change of 49.2 J /kg K in a field change from 0 to 5 T at 253 K, which is superior to the previous studies [12,15].

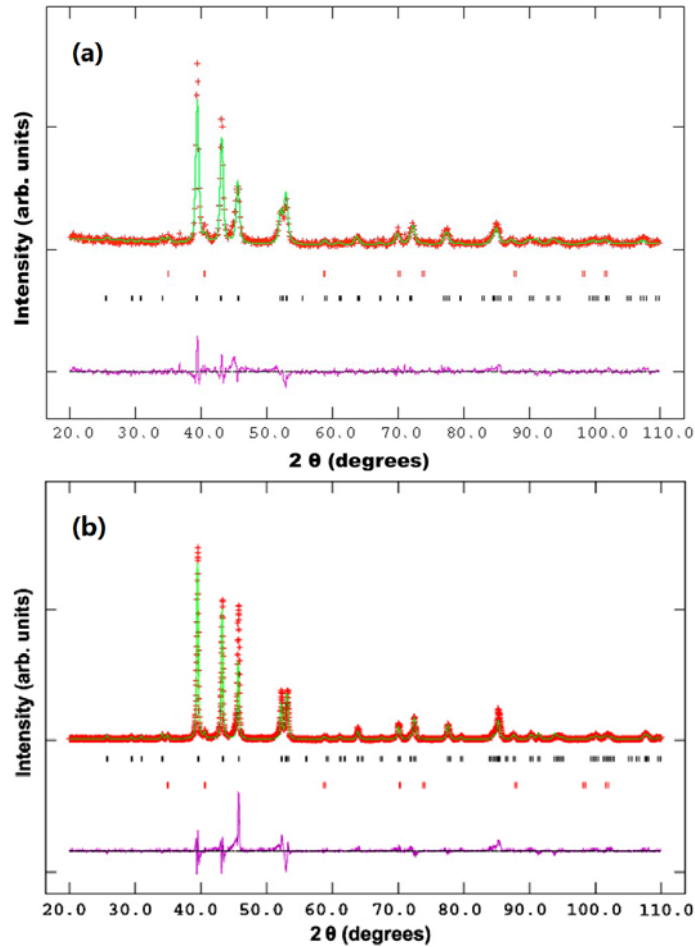

**Fig. 1.** (a) and (b) are the observed and calculated XRD patterns of bulk $Mn_{1.1}Fe_{0.9}P_{0.8}Ge_{0.2}$ alloy. The bottom curve is the difference between the observed and the calculated intensities. The black and red rows of vertical bars indicate the Bragg reflection positions of MnFePGe and MnO compounds, respectively[17,18].

To make the reaction among elemental powders more adequately, a high energy ball milling process was added before SPS[18]. In detail, the mixed elemental powders were ball milled under an argon atmosphere for 1.5 h prior to SPS. From Fig. 1(b) we can find the final $Mn_{1.1}Fe_{0.9}P_{0.8}Ge_{0.2}$ sample also composed of main $Fe_2P$ type phase and a minor impurity phase of MnO. On the other hand, magnetic measurements show that the new sample bears a remarkable enhanced maximum magnetic entropy change of 61.8 J /kg K in a field change from 0 to 5 T at its $T_C$ of 251 K, which is more or less equal to the sample without ball milling. Furthermore, it is found that the thermal hysteresis of the new sample is also increased by 10 K. It is therefore concluded that the ball milling process plays an important role to the modification of the MCE of the MnFe(PGe) compounds.

Upon our later investigation with neutron powder diffraction technique, we find that the chemical homogeneity and grain size of the MnFe(PGe) compounds have substantial influence to their MCE[19]. Therefore, a heat treatment process was added to the SPS sample[20]. In detail, the as-sintered samples were solution treated at 950 °C for 15 h, followed by annealing at 800, 850, 900, and 950 °C for 48 h, respectively, before they were quenched into ice water. Fig. 2 shows the magnetic entropy change of as-sintered and annealed $Mn_{1.1}Fe_{0.9}P_{0.8}Ge_{0.2}$ samples in a field change of 0-3 T. The annealed sample exhibits increased $T_C$ and remarkably enhanced magnetic entropy change compared with those of as-sintered one. Moreover, the thermal hysteresis reduces from 15 K for as-sintered sample to 9 K for annealed sample, indicating the key role of proper preparation to the magneto-caloric properties of the MnFe(PGe) compounds.

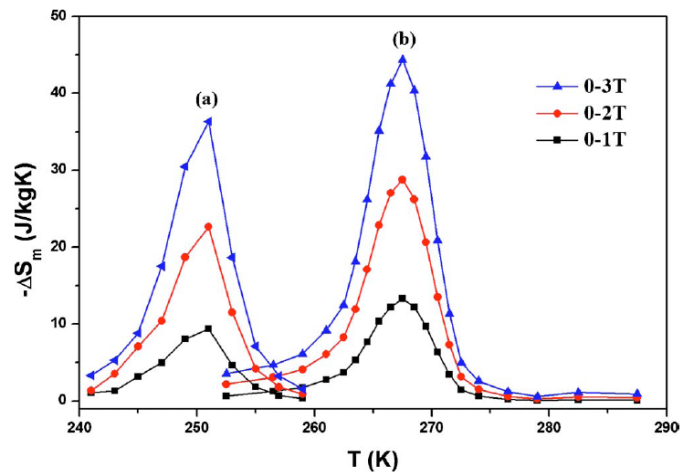

**Fig. 2** Temperature dependence of the magnetic entropy change of the bulk $Mn_{1.1}Fe_{0.9}P_{0.8}Ge_{0.2}$ alloys measured in a magnetic field change from 0 to 1, 2, and 3 T.[20]

Table 1. Structural parameters of $Mn_{1.1}Fe_{0.9}P_{0.8}Ge_{0.2}$ at 295 and 10 K. Space group *P-62m*. Atomic positions: Mn:3g(x,0,1/2); Fe/Mn:3f(x,0,0);P/Ge(1):1b(0,0,1/2); P/Ge(2):2c(1/3,2/3,0).[21]

| Atom | Parameters | 295 K PMP | 10 K FMP |
|---|---|---|---|
|  | a (Å) | 6.06137(7) | 6.17811(9) |
|  | c (Å) | 3.46023(5) | 3.30669(7) |
|  | V (Å) | 110.098(3) | 109.304(3) |
| Mn | x | 0.5916(3) | 0.5956(5) |
|  | B (Å$^2$) | 0.77 (2) | 0.58 (2) |
|  | M ($\mu_B$) |  | 3.0(1) |
|  | n (Mn/Fe) | 0.998/0.002(3) | 0.988/0.012(4) |
| Fe/Mn | x | 0.2527(1) | 0.2558(2) |
|  | B (Å$^2$) | 0.77(2) | 0.58(2) |
|  | M ($\mu_B$) |  | 1.7(1) |
|  | n (Fe/Mn) | 0.928/0.072(3) | 0.922/0.078(4) |
| P/Ge(1) | B (Å$^2$) | 0.55(4) | 0.54(4) |
|  | n (P/Ge) | 0.947/0.053(8) | 0.93/0.07(1) |
| P/Ge(2) | B (Å$^2$) | 0.55(4) | 0.54(4) |
|  | n (P/Ge) | 0.726/0.274(4) | 0.736/0.264(6) |
| $R_P$(%) |  | 5.25 | 7.05 |
| $wR_P$(%) |  | 6.65 | 8.75 |
| $\chi 2$ |  | 1.276 | 1.913 |

## 3. Structure and phase transformation of MnFe(PGe) compounds

### 3.1 Crystal and magnetic structures of MnFe(PGe) compounds

Liu et al[21] reported crystal structure and magnetic structure of $Mn_{1.1}Fe_{0.9}P_{0.8}Ge_{0.2}$ compound investigated by Neutron diffraction technology. As shown in Table 1, the compound bearing $Fe_2P$-type hexagonal structure with space group of $P\bar{6}2m$ is single paramagnetic phase (PM) above 255 K and pure ferromagnetic phase (FM) below 10 K, respectively, and FM and PM phases possess obviously different lattice parameters, i.e. the *a* axis is 1.3% longer and the *c* axis is 2.6% shorter in the FM phase compared to the PM phase. Fig. 3 shows the sketch of the crystal structure and magnetic structure of the $Mn_{1.1}Fe_{0.9}P_{0.8}Ge_{0.2}$ compound. The Mn atoms are coplanar with the

P/Ge(1) atoms and the Fe/Mn atoms are coplanar with P/Ge(2). The intra-plane transition metals form a triangular configuration. The Mn atoms are surrounded by four P/Ge(2) atoms located on the layers above and below and by one apical P/Ge(1) atom on the same layer, forming a pyramid. The Fe/Mn site is coordinated by two P/Ge(2) atoms located on the same layer and two P/Ge(1) atoms in the layers above and below, forming a tetrahedron. It is observed that the the $3g$ sites are completely occupied by Mn atoms sharing the same plane with P/Ge(1) atoms on $1b$ sites, while the $3f$ sites has 93% Fe and 7% Mn distributed randomly, which is in the same plane with the P/Ge(2) atoms on $2c$ sites. Different from the total disordering in MnFePAs compounds, there is a small degree of site preference for Ge atoms on $2c$ and $1b$ sites.[22]

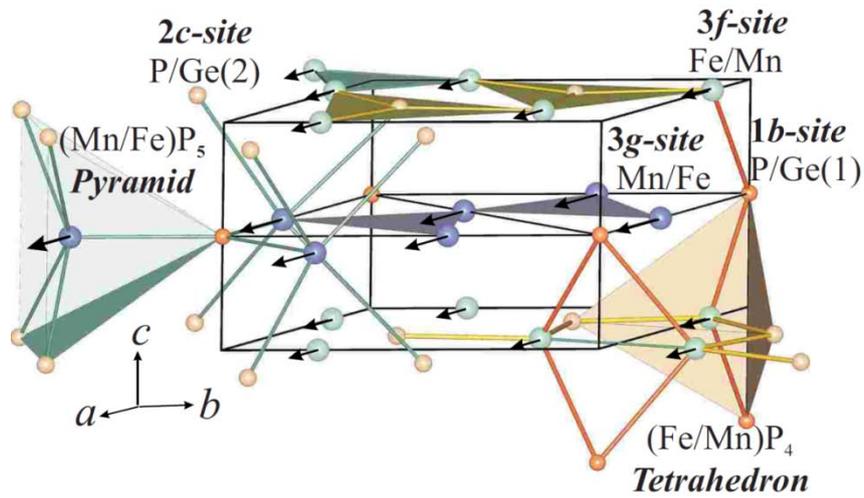

Fig. 3 Crystal structure and magnetic structure of $Mn_{1.1}Fe_{0.9}P_{0.8}Ge_{0.2}$ alloy.[19]

The magnetic structure of $Mn_{1.1}Fe_{0.9}P_{0.8}Ge_{0.2}$ compound has been determined as a *P11m* magnetic symmetry, in which Mn and Fe moments are parallel to each other in the *a-b* plane. This is not like it in MnFePAs compounds which have them lie in the *a-c* plane or along the *c* axis [22]. The refined moments for Mn atoms on $3g$ sites and Fe/Mn atoms on $3f$ sites at 245 K are 2.9 and $0.9\mu_B$, respectively. These results are similar to the situations in other compounds with the $Fe_2P$-type of structure, where the $3g$ site has a larger moment than it on $3f$ site [14,22]. From Table 2, it can be seen that the inter-atomic distances between Fe/Mn atoms and the surrounding P/Ge atoms in the pyramid and tetrahedron coordination are very different, which will strongly affect the bonding strength between them and thus induce different magnetic moments.

Table 2. Selected inter-atomic distances (Å) at 295 and 10K.[21]

|  | 295 K | 10 K |
|---|---|---|
| | Intraplane metal to metal | |
| Mn-Mn | 3.180(1) | 3.254(2) |
| Fe/Mn-Fe/Mn | 2.653(2) | 2.738(2) |
| | Interplane metal to metal | |
| Mn-Fe/Mn | 2.686(2) | 2.672(3) |
| Mn-Fe/Mn | 2.771(1) | 2.743(2) |
| | Fe/MnP$_4$ tetrahedron | |
| Fe/Mn-P/Ge(2) *2 | 2.3109(6) | 2.3358(7) |
| Fe/Mn-P/Ge(1) *2 | 2.3039(6) | 2.2874(8) |
| | MnP$_5$ pyramid | |
| Mn-P/Ge(1) | 2.476(2) | 2.499(3) |
| Mn-P/Ge(2) *4 | 2.5225(5) | 2.5026(7) |

*3.2 Crystal structure evolution during the phase transition in MnFe(PGe) compounds*

It is known from Table 1 that during the temperature induced PM↔FM phase transition in Mn$_{1.1}$Fe$_{0.9}$P$_{0.8}$Ge$_{0.2}$ compound, the *c* axis of the crystal lattice will be shortened while the *ab* plane will be expanded. Moreover, more specific and unexpected bonding characteristics of this material during the phase transition have been discovered by a detailed inspection of the structure variation [21,23].

Fig. 4(a) shows the temperature variations in the lattice parameters for the PMP and FMP of Mn$_{1.1}$Fe$_{0.9}$P$_{0.8}$Ge$_{0.2}$ compound. The *a*-axis lattice parameter increases and the *c*-axis lattice parameter decreases abruptly at the transition. However, aside from the sharp changes that occur at the phase transition, there is little variation with temperature. The temperature dependence of the relevant metal-metal bond distances in or between the neighbor layers is shown in Fig. 4(b). During the transition from PMP to FMP, the intra-layer metal-metal bond distances show a significantly increase, while the interlayer distances either remain almost constant or decrease slightly, indicating that the shortening of the *c* axis is mainly due to the decrease of the

P/Ge(1)-Fe/Mn-P/Ge(1) angle.

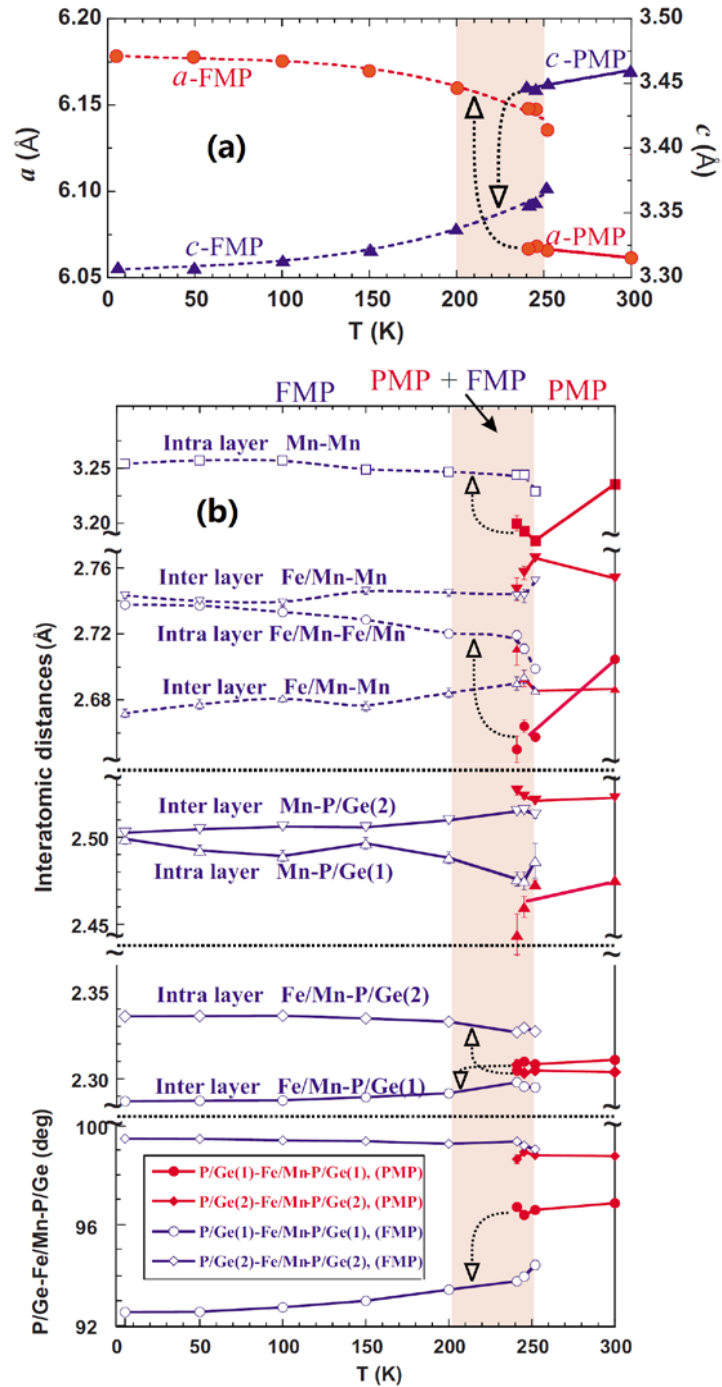

Fig. 4 Variation in lattice parameters (a) and the metal-metal bond distances (b) as function of temperature. The changes occurring at the transition are indicated by the arrows.[21]

Figure 5 shows the relative atomic positions in the $a$-$b$ plane in the PM and FM phases, and the atomic shifts between the two phases are indicated by the arrows. On the $z = 0$ layer, rotations of the P/Ge(2) atoms result in the substantial increase of Fe/Mn-Fe/Mn distances and

Fe/Mn-P/Ge(2) distances. On the $z=1/2$ layer, the same behavior occurs for the Mn-Mn distances, with only a slight variation in the Mn-P/Ge(1) separations. It is therefore found that only the triangular framework of magnetic atoms has been mostly affected during the phase transition, exhibiting that structural and magnetic transitions are tightly connected to each other and hence any modulating of inter-atomic distances will directly affects the transition temperature and thus the MCE properties.

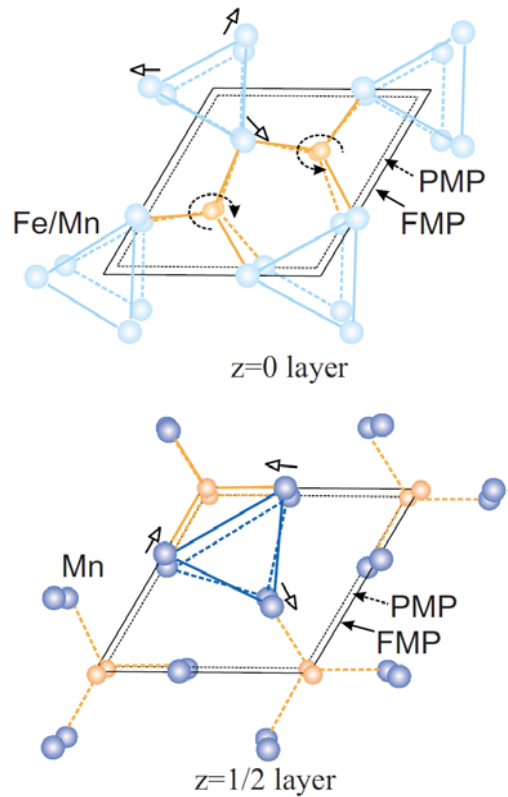

Fig.5 Projections along the $c$ axis of the atomic arrangement in the $z=0$ and $z=1/2$ layers of the structure. The atomic shifts and the rotations about the P/Ge atoms taking place at the transition are indicated by the arrows. The outlines of the unit cell and the bonds between the atoms are shown by continuous and broken lines for the FMP and PMP, respectively.[21]

According to the structure variation during the phase transition, one can easily image the structure evolution during phase transition under a certain pressure since the expansion of crystal lattice will be preferred by the formation of FM phase. Consequently, in an opposite way, application of pressure should inhibit the formation of the FM phase and thus decrease the $T_C$. This is the situation that had been observed by Liu et al [21] when they applied a pressure up to

1GPa on the sample. As shown in Fig. 6, at 245 K, the phase fraction of PMP continuously increases while that of FMP reduces correspondingly when the applied pressure is increased from 0 to 0. 92 GPa.

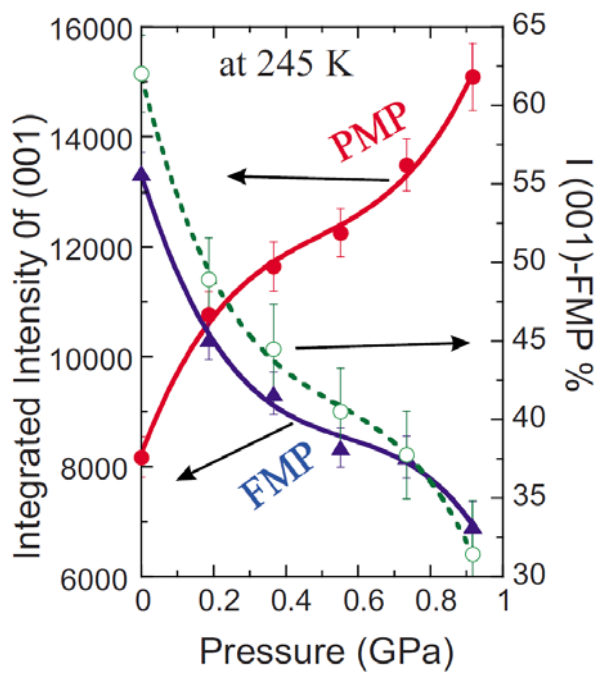

Fig. 6 Integrated intensities for the PM and FMreflections as a function of pressure, showing that the intensity of (001)-PM peak increases and that of the (001)-FM peak decreases. The(green) broken line shows the relative intensities of the two peaks.[21]

Fig. 7(a) shows the magnetic field variations in the lattice parameters for the paramagnetic phase (PMP) and ferromagnetic phase (FMP) of $Mn_{1.1}Fe_{0.9}P_{0.8}Ge_{0.2}$ compound. As the field strength increases, the *a*-axis lattice parameter increases and the *c*-axis lattice parameter decreases simultaneously in FMP. On the contrary, the lattice parameters in PMP show an opposite variation, indicating its chemical composition is different from that of FMP. The magnetic field dependence of the relevant metal-metal bond distances in or between the neighbor layers is shown in Fig. 4(b). In going from the PMP to the FMP, the variation in the metal-metal distances is positive and large for the intra-layer bonds and rather small and negative in the case of the interlayer bonds. It is worth to note that all the distances remain remarkably constant when the field varies, indicating that the crystal and magnetic structures of the PMP and FMP do not change markedly during the transition. Moreover, variations of lattice parameters with deceasing temperature or increasing field are similar, demonstrating that the effect of temperature on the nature of the transition is basically equivalent to that of an applied magnetic field.

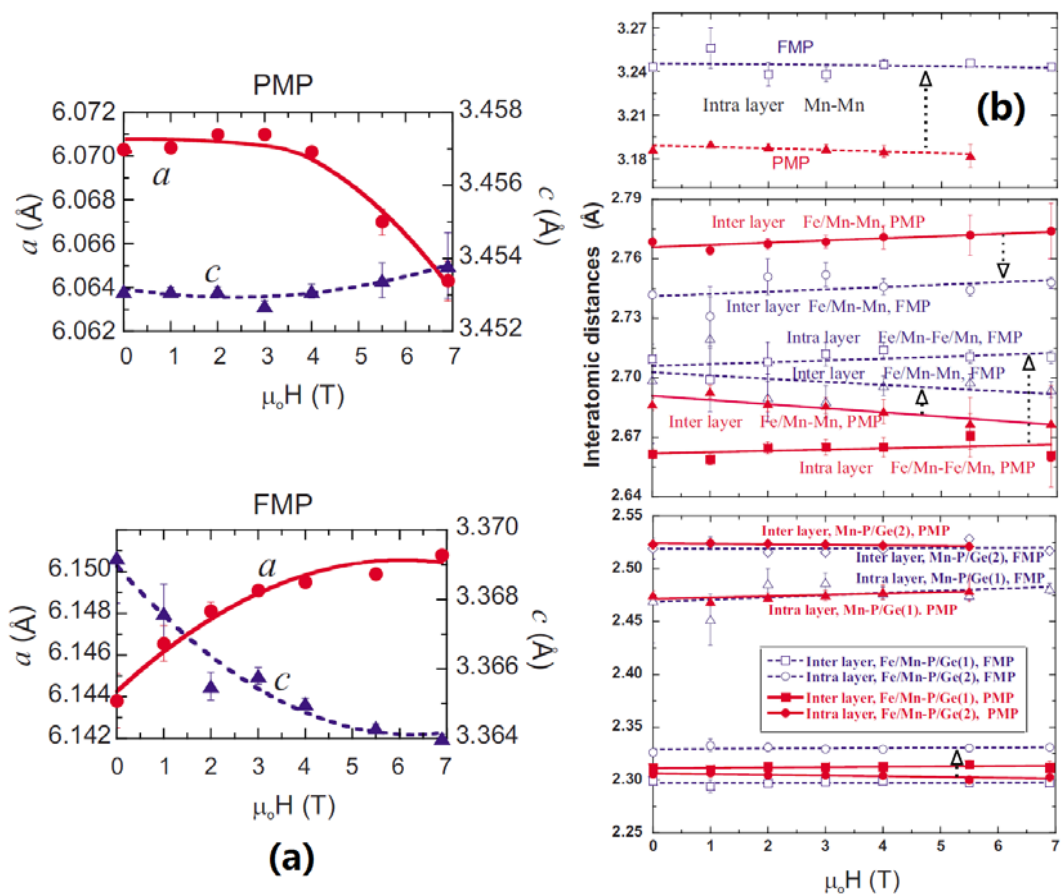

Fig. 7 (a) The magnetic field variations in the lattice parameters for the paramagnetic phase (PMP) and ferromagnetic phase (FMP) of $Mn_{1.1}Fe_{0.9}P_{0.8}Ge_{0.2}$ compound; (b) Bond distances at 255 K as a function of applied magnetic field. The intralayer metal-metal distances show sharp increases in going from the PMP to the FMP while the interlayer distances decrease (these variations are indicated by the arrows in the figure). The bond distances of the metal—P/Ge do not show a significant variation with field.[21]

*3.3 Correlation between phase transition and MCE in MnFe(PGe) compounds*

The NPD experiments release an important knowledge about this material: the magnetic field or temperature induced magnetic-entropy change is directly controlled by the fraction of PMP and FMP during the first-order transition. As shown in Fig.8 (a), the refined phase fraction from NPD data reveal that the FM-PM transition is almost fully accomplished when temperature goes up to above 255 K. However, during the PM-FM transition, there will always be about 4.5% of the sample remains in PM state until the temperature decreases below 10 K. The same situation happens during the first-order structural transition driven by external magnetic field, as shown Fig.8 (b), only 70% of PM phase was converted into the FM phase in a field of 5 T. Liu et al [19] has normalized the magnetic entropy change $\Delta S_m$) to the magnetization, [001]-FM intensity and fraction of FM phase, as shown in Fig. 8(c) and 8(d). The excellent agreement between these parameters and the linear relationship between $\Delta S_m$ and the FM-phase fraction strongly suggest that FM-phase fraction has direct connection to the magnetocaloric effect.

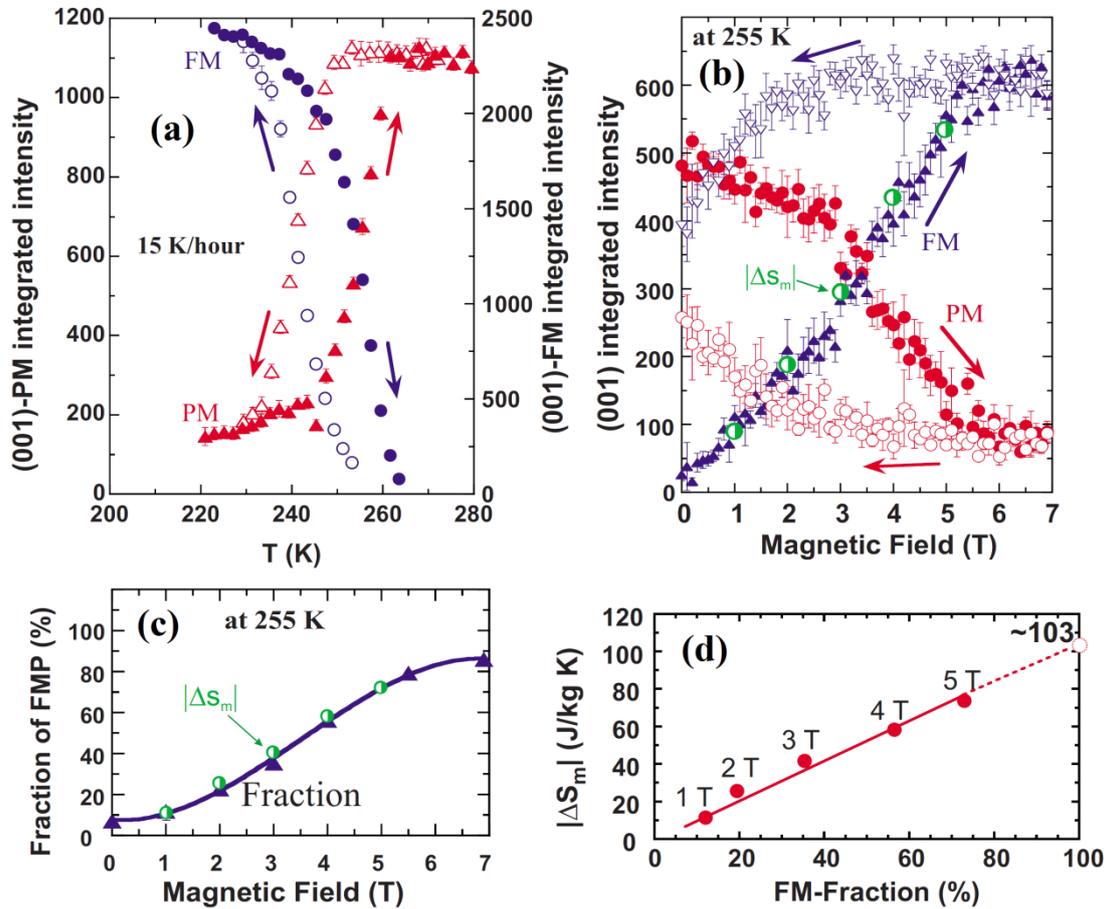

Fig. 8.(a) Integrated intensities of the (001) reflections for the PM and FM phases as a function of temperature on cooling and warming;(b) Field dependence of the integrated intensities of the (001) reflections for the PM phase and FM phase at 255 K, showing that the FM phase fraction tracks the magnetization data. For comparison, data normalized from the magnetic-entropy change $\Delta S_m$ are also shown; (c) Fraction of the ferromagnetic phase (FMP) at 255 K as the field increases. The FMP fraction increases smoothly to ~86%. Data normalized from $\Delta S_m$ are shown for comparison; (d)$\Delta S_m$ as a function of the ferromagnetic phase fraction. The linear relationship shows that the entropy change simply tracks the FM-phase fraction. $\Delta S_m$ is projected to be ~103 J /kg K if the transition went to completion for this sample.[19]

## 4. Evaluation of MCE of MnFe(PGe) compounds

The magnetic entropy change ($\Delta S_m$) is the key parameter to evaluate the MCE of magnetic refrigerant materials. A popular way to obtain $\Delta S_m$ is from the isothermal magnetization curves with either Maxwell [24] or Clausius-Clapeyron relation [25]. However, this method is under debate in application to first order magnetic transition [8,26-28]. Another way is to derive $\Delta S_m$ from heat capacity measurements under different magnetic fields. However, the conventional methods of measuring heat capacity are time consuming and not suitable for materials with first order transition since a heat input does not necessarily lead to a temperature modification in the sample due to the latent heat. It is therefore necessary to explore new method for evaluation of MCE of the magneto-caloric materials with first order magnetic transition.

A reliable method to obtain entropy change ($\Delta S$) is the differential scanning calorimeter (DSC) which measures the heat flux, while the temperature of the calorimeter is continuously changed, and with proper integration of the calibrated signal the latent heat of the transition can be obtained. Therefore DSC is particularly suited in the case of first order phase transitions since it yields both the latent heat and the entropy changes associated with the transitions. Moreover, DSC measurements directly provide both the magnetic and the structural contributions to $\Delta S$ [29].

The results discussed in section 3 suggests that the effect of temperature on the nature of the transition is basically equivalent to that of an applied magnetic field, demonstrating DSC measurements a potential way to determine the MCE of the magneto-caloric materials. To clarify the relationship between $\Delta S_m$ obtained from magnetic measurements and $\Delta S$ derived from DSC measurements, Yue et al[29] investigated the structure evolution, magnetic transition, as well as the MCE of $Mn_{1.1}Fe_{0.9}P_{0.76}Ge_{0.24}$ compound.

Fig. 9(a) shows the PM phase fraction as a function of temperature during the magnetic transition in $Mn_{1.1}Fe_{0.9}P_{0.76}Ge_{0.24}$ compound. The compound is in the fully PM state above 272 K, and below this point the PM and FM phases coexist. It is found that 85.5% of the sample transforms quickly between 272 and 263 K, but the rest of the compound remains in the PM state and slowly transform into FM state with decreasing temperature. Fig. 9(b) shows the PM phase fraction as a function of the magnetic field in the compound at 272 K. It is observed that most of the PM phase transform into the FM phase between 1.4 and 4.1 T, and the transition is not completed even when the field strength reaches 7T, with 17% of the sample remaining in the PM state. Note that the behaviors of the transition with temperature and with magnetic field are quite similar to each other.

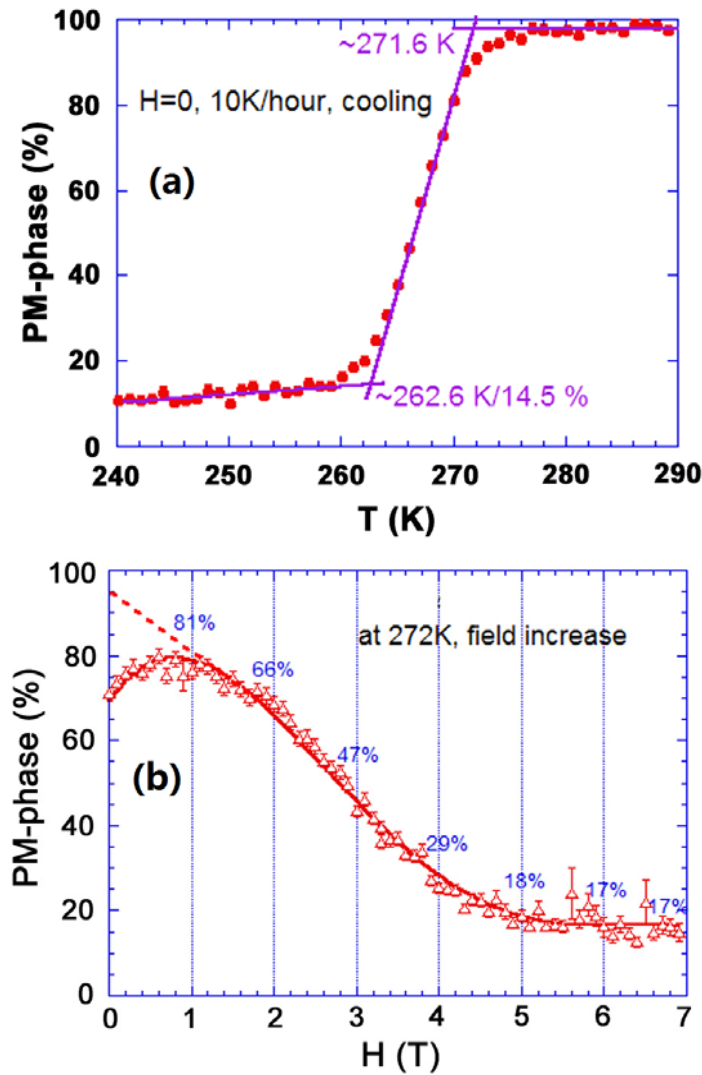

Fig. (a) PM phase fraction as a function of temperature. About 85.5% of the paramagnetic phase transforms to the ferromagnetic phase in the temperature interval from 272 to 263 K, while the remaining 14.5% changes only very slowly below 263K with further decrease of temperature; (b) PM phase fraction as a function of magnetic field. About 82% of the paramagnetic phase changes quickly to the ferromagnetic phase from 0 to 5 T, while the remaining 18% changes only slowly with increased magnetic field.[29]

Table 3 structure parameters and magnetic moments of FMP in Mn$_{1.1}$Fe$_{0.9}$P$_{0.76}$Ge$_{0.24}$ compound at 259K/0T and 271K/3T.

| | | 295 K/0 T | | 271 K/3 T | |
|---|---|---|---|---|---|
| Parameters | | PMP | FMP | PMP | FMP |
| | Phase Fraction | 14.2% | 81.7% | 16.6% | 79.9% |
| | $a$ (Å) | 6.0799(6) | 6.15891(6) | 6.0775(5) | 6.15511(8) |
| | $c$ (Å) | 3.4463(5) | 3.35934(6) | 3.4505(5) | 3.3652(1) |
| | $V$ (Å) | 110.32(2) | 110.355(3) | 110.37(2) | 110.41(2) |
| Mn | $x$ | 0.594137 | 0.5961(4) | 0.5941 | 0.5943(6) |
| | $B$ (Å$^2$) | 0.11(6) | | 0.49(6) | |
| | $M$ ($\mu_B$) | | 3.49(7) | | 3.4(1) |
| | n (Mn/Fe) | 0.991/0.009 | 0.974/0.026(4) | 0.991/0.009 | 0.974/0.026 |
| Fe/Mn | $x$ | 0.254921 | 0.2543(1) | 0.2549 | 0.2550(2) |
| | $B$ (Å$^2$) | 0.57(2) | | 0.92(3) | |
| | $M$ ($\mu_B$) | | 0.71(8) | | 0.6(1) |
| | n (Fe/Mn) | 0.939/0.061 | 0.95/0.05 | 0.939/0.061 | 0.95/0.05 |
| P/Ge(1) | $B$ (Å$^2$) | 0.66(6) | | 1.00(6) | |
| | n (P/Ge) | 0.906/0.094 | 0.78/0.22(2) | 0.906/0.094 | 0.78/0.22 |
| P/Ge(2) | $B$ (Å$^2$) | 0.33(4) | | 0.63(4) | |
| | n (P/Ge) | 0.712/0.288 | 0.68/0.32(2) | 0.712/0.288 | 0.681/0.319 |
| | | Intra plane metal to metal distance | | | |
| Mn-Mn | | 3.197(6) | 3.245(1) | 3.196(3) | 3.238(2) |
| Fe/Mn-Fe/Mn | | 2.6845(2) | 2.713(2) | 2.6835(2) | 2.719(3) |
| | | Inter plane metal to metal distance | | | |
| Mn- Fe/Mn | | 2.6875(2) | 2.693(2) | 2.6882(2) | 2.683(3) |
| Mn- Fe/Mn | | 2.7633(2) | 2.751(1) | 2.7640(2) | 2.758(2) |
| | | Fe/MnP4 tetrahedron distance | | | |
| Fe/Mn-P/Ge(2) ×2 | | 2.3023(2) | 2.3346(6) | 2.3014(2) | 2.3304(9) |
| Fe/Mn-P/Ge(1) ×2 | | 2.3176(2) | 2.2968(6) | 2.3188(2) | 2.301(1) |
| | | MnP5 pyramid distance | | | |

| | | | | |
|---|---|---|---|---|
| Mn-P/Ge(1) | 2.4676(2) | 2.487(2) | 2.4667(2) | 2.496(4) |
| Mn-P/Ge(2) ×4 | 2.5253(2) | 2.5165(5) | 2.5262(2) | 2.5152(9) |
| | Intra plane angle | | | |
| P/Ge(1)-Fe/Mn-PGe(1) | 96.06(1) | 94.00(3) | 96.15(1) | 93.98(5) |
| P/Ge(2)-Fe/Mn-PGe(2) | 99.339(0) | 99.21(3) | 99.399(0) | 99.36(5) |
| $R_P$(%) | 4.66 | | 8.11 | |
| $wR_P$(%) | 6.08 | | 10.03 | |
| $\chi^2$ | 3.358 | | 2.136 | |

To clarify the evolution of crystal structure and magnetic structure of the PMP and FMP in the $Mn_{1.1}Fe_{0.9}P_{0.76}Ge_{0.24}$ compound during transition process, the structure parameters as well as magnetic moments of the compound at four temperature-field values, 267K/0T, 259K/0T, 271K/3T, and 271K/6.9T, were investigated. It is interesting to found that the phase fractions, lattice parameters, atomic occupancy factors, bond distances and angles, and magnetic moments of the FMP in the compound at 259K/0T and 271K/3T are almost same to each other within uncertainties, as shown in Table 3. However, if the magnetic field increases to 6.9 T at 271 K, the magnetic moments will increase by 15%. It is therefore concluded that the magnetic and structural transitions in $Mn_{1.1}Fe_{0.9}P_{0.76}Ge_{0.24}$ compound induced by temperature are almost identical to the transition induced by the application of a magnetic field. Under these situations, it is expected that the entropy changes are also approximately the same in the two cases.

Fig. 10 shows the temperature dependence of the entropy upon cooling and warming. The transition started at 271 K and ended at 261 K under cooling, and started at 282 K and ended at 273 K under warming, exhibiting a ~11K thermal hysteresis. The average integrated entropy changes from 240K to 300K upon cooling and from 261K to 300K upon warming are 33.81 and 30.76 J/kg•K, respectively. Note that only 81.7% of the PMP was converted into FMP at 259 K, and there is still 4% MnO impurity in the sample. Hence average entropy changes as high as 42.71 J/ kg•K could be achieved if the phase transformation goes to completion in a pure sample.

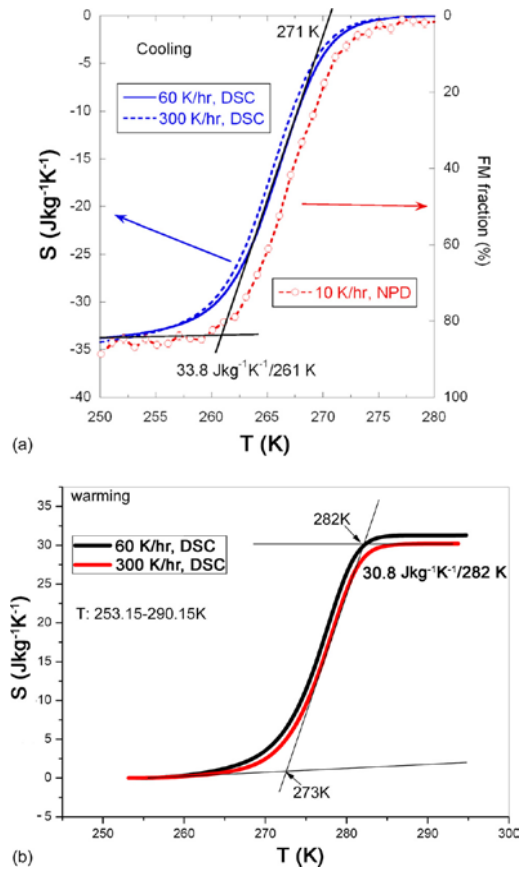

Fig. 10 PM-FM phases transformation entropy of $Mn_{1.1}Fe_{0.9}P_{0.76}Ge_{0.24}$ as a function of temperature calculated from DSC data, in rates of 1 K/min or 5 K/min. (a) cooling, with DS¼33.8 J/Kg K, and (b) warming, with DS¼30.8 J/Kg K.[29]

Fig. 11 shows the temperature dependence of the $\Delta S_m$ in the $Mn_{1.1}Fe_{0.9}P_{0.76}Ge_{0.24}$ compound obtained using magnetization curves and the Maxwell relation. As shown in the figure, the measurements were made via four modes: (i) increasing temperature increasing field; (ii) increasing temperature-decreasing field; (iii) decreasing temperature-increasing field; (iv) decreasing temperature-decreasing field. The results from the four types of measurements give similar values of $\Delta S_m$. Note that the four types of measurements result different magnetic transition process in the compound, so an average $\Delta S_m$ value of 46.5 J/kg•K for fields from 0 to 5T was obtained, which translates to maximum value of 58.1 J/kg•K after correction for the impurity and phase fraction.

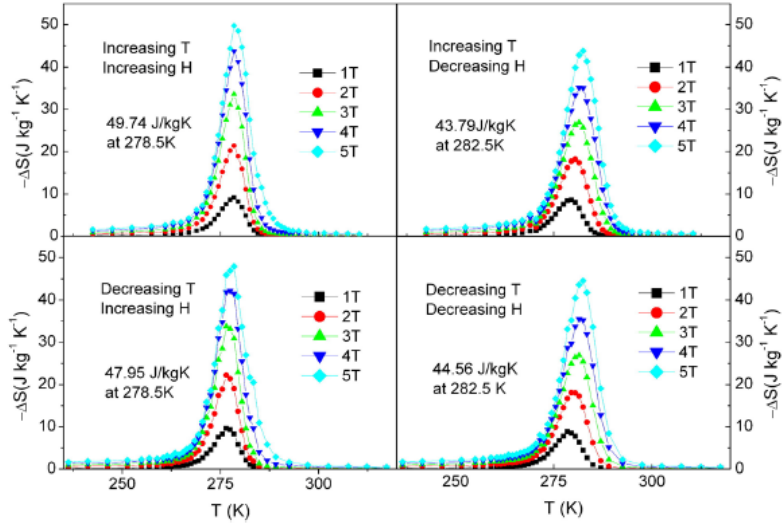

Fig. 11. Temperature dependence of the magnetic entropy change in bulk $Mn_{1.1}Fe_{0.9}P_{0.76}Ge_{0.24}$ as a function of magnetic field up to 5 T determined using the Maxwell relation. The steps of temperature and magnetic field used in the measurements were 1K and 0.1 T.[29]

It is easy to find that the $\Delta S_m$ value via magnetic measurement is substantially larger than the measured $\Delta S$ obtained directly from the DSC technique. However, based on above investigation of the lattice parameters and magnetic moments of the $Mn_{1.1}Fe_{0.9}P_{0.76}Ge_{0.24}$ compound at different temperature/field situations, we may need to use an average $\Delta S_m$ value for fields from 0 to 3T instead, which is more or less equal to the $\Delta S$ from DSC measurements. It is therefore concluded that detailed comparisons are only reliable using the DSC technique, on well characterized samples where the structures and phase fractions are known.

5. Origin of virgin effect in MnFe(PGe) compounds

The "virgin effect" is a new phenomenon firstly observed in $Fe_2P$-type $Mn_{2-y}Fe_yP_{1-x}T_x$ (T= As, Ge, and Si) compound[14,30,31]. Further investigations show that this new phenomenon also occurs in other Mn-based compound like MnAs [32]. A typical "virgin effect" in $Mn_{1.1}Fe_{0.9}P_{0.8}Ge_{0.2}$ compound was shown in Fig. 12. The as-prepared sample shows a significantly lower transition temperature on first cooling than after it has undergone thermal cycling down to 50 K [30]. In detail, with decreasing temperature, a PM-FM transition starts at about 200 K for the as-prepared sample and ends at about 100 K. These results imply that this magnetic transition may co-occur with a structural change and may need a very large overcooling as driving force. After it has undergone thermal cycling to 20 K, the FM-PM transition will be completed at about 240 K during the subsequent warming process. For the second thermal cycle, the PM-FM transition starts at about 230 K during cooling process while FM-PM transition ends at 240 K during heating process. Further thermal cycles have no effect on $T_C$.

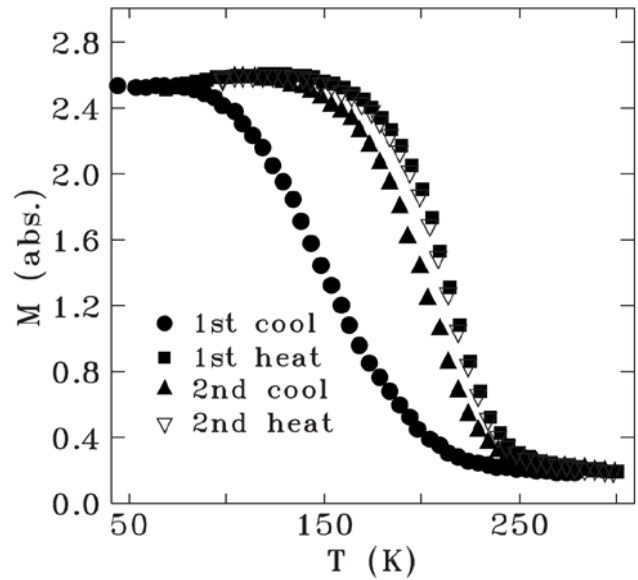

Fig. 12 Temperature dependence of magnetization, $MT$, for $Mn_{1.1}Fe_{0.9}P_{0.8}Ge_{0.2}$ under a magnetic field of 10 mT.

Though experimental results show that virgin effect happens only during the first several thermal cycles and will has no influence on the MCE performance of the materials, it is still important to explore its origin and relation to the magneto-structural change during the first-order transition. According to some neutron diffraction results of MnFePSi compounds, the virgin effect is thought to be related to the intrinsic strain effect in the samples, which has been proved not very precise by a recent research.

Liu et al has done a thoroughly research on the virgin effect in $Mn_{1.1}Fe_{0.9}P_{0.8}Ge_{0.2}$ by NPD and first-principles calculations[33]. Firstly, they found from the refined NPD data that there is an irreversible structural change during the first thermal cycle, i.e. a structural relaxation accompanying the magneto-structural transition. According to their statement, since the sample is prepared via the non-equilibrium processes like ball milling and SPS, there will be many meta-stable distorted structural units with random distribution of P/Ge atoms on both $1b$ and $2c$ sites. Due to the atomic size difference between Ge and P, the Fe-centered tetrahedron and Mn-centered pyramid will be distorted and cause local fluctuation of the inter-atomic distances and thus the magnetic properties.

To prove this point the inter-atomic exchange coupling parameters for the paramagnetic and ferromagnetic phases of the material has been carried out. It turns out that the magnetic exchange interaction is highly sensitive to the inter-atomic distances between magnetic atoms (Fe,Mn). The existence of the structural distortion results in the local fluctuation of the inter-atomic distances and the related magnetic exchange interactions. Therefore, here comes the scenario of this physic effect. During the first cooling, the local fluctuation of magnetic exchange interaction makes the structural change around $T_C$ more difficult to happen. Once the transition is accomplished, however, the structure will have the opportunity to relax to a more stable configuration under the motivation of magnetic exchange interactions. When the temperature warms up, the atoms therefore not necessarily restore into the same atomic positions in the as-prepared sample before the thermal cycle, and thus cause significant increase of the $T_C$. Further thermal cycles will not be affected by this effect because most of the structural relaxation is complete during the first thermal cycle.

## 6. Summary

We have prepared MnFe(PGe)-based compounds by using ball milling, SPS, and annealing successively. Structure evolution and phase transformation in the compounds as a function of temperature, pressure, and magnetic field were reported. In this paper, correlations among preparation conditions, magnetic and crystal structures, and magnetocaloric effects (MCE) of the MnFePGe-based compounds are reviewed.

(1) Due to the unique sintering mechanism, the spark plasma sintering technique is an effective way to prepare $Fe_2P$ type MnFe(PGe)-based compounds with high purity. To achieve chemical composition and microstructure homogeneity in the compounds, a subsequent annealing process is necessary.

(2) The crystal structure and magnetic structure of the $Mn_{1.1}Fe_{0.9}P_{0.8}Ge_{0.2}$ compound has been determined as $Fe_2P$-type hexagonal structure (space group of $P\bar{6}2m$) and $P11m$ magnetic symmetry, respectively. The PMP-FMP structural and magnetic transition can be facilitated by temperature, pressure, and magnetic field via modulating of inter-atomic distances in the compounds. Moreover, the FM-phase fraction in the compounds during PMP-FMP transition has direct connection to the MCE of the compounds.

(3) Reliable entropy changes and other important thermodynamic properties associated the transitions can be obtained by DSC measurements combined with the structure and magnetic properties of the MnFe(PGe)-based compounds being investigated, providing an effective and easy way to evaluate the MCE of the compounds.

(4) The "virgin effect" related irreversible structural change in MnFe(PGe)-based compounds results from the structural relaxation of metastable structural distortions in the as-prepared state upon thermal cycling. This behavior originates from the interplay between structural distortion and inter-atomic exchange interaction.